\documentstyle[11pt,twoside,graphicx,jltp]{article}

\title{Physics of Proteins at Low Temperature}

\author{Vladimir V. Ponkratov, Josef Friedrich\address{Physics Department E14 and Lehrstuhl f\"ur Physik Weihenstephan, Technische Universit\"at M\"unchen, D-85350 Freising, Germany}, Jane M. Vanderkooi\address{Department of Biochemistry and Biophysics, University of Pennsylvania, Philadelphia, PA 19104, USA}, Alexander L. Burin\address{Department of Chemistry, Tulane University, New Orleans, LA 70118, USA} and Yuri A. Berlin\address{Department of Chemistry, Northwestern University, IL 60208, USA}}

\runninghead{\bf V.V.~Ponkratov {\boldmath $et\,al.$},}{Physics of Proteins at Low Temperature}

\begin{document}

\begin{abstract}
We present results of a hole burning study with thermal cycling and waiting time spectral diffusion experiments on a modified cytochrome - c protein in its native as well as in its denatured state. The experiments show features which seem to be characteristic for the protein state of matter and its associated dynamics at low temperature. The properties responsible for the observed patterns are organisation paired with randomness and, in addition, the finite size which gives rise to surface and solvent effects. We discuss some general model approaches which might serve as guide lines for understanding these features.

\end{abstract}

\maketitle

\section{INTRODUCTION}
One of the most fascinating aspect of protein physics is the interplay between structural order and randomness.\cite{Frauen_1998,Frauen_1994} Order is reflected in the high structural organisation which, in turn, shows up in rather well resolved diffraction patterns. Structural order and organization are prerequisites for a proper functioning. However, also randomness seems to be a prerequisite for folding and functioning.\cite{Dill_1997} Randomness is reflected in many spectroscopic and kinetic properties of proteins. For instance, inhomogeneous broadening of optical absorption or emission bands is a direct manifestation of structural heterogeneity in the environment of the chromophore. The corresponding widths in proteins are comparable to the widths observed for chromophores in glasses. Another example of structural heterogeneity concerns the mean square displacement ${<x^2>}$ of the aminoacid residues, which can be measured in an X-ray experiment. The corresponding values at low temperature are not determined by zero point quantum motions but, instead, are much larger due to conformational disorder.\cite{Frauen_1979,Hartmann_1982} Kinetic experiments, too, are very sensitive to the presence of conformational disorder. For instance the rebinding of CO in myoglobin at low temperature, is characterized by an extremely strong dispersion of the rate parameters which, in turn, is attributed to a dispersion of activation barriers.\cite{Frauen_1991,Berlin_1995} It is obvious that this dispersion of barriers has its roots in a dispersion of structures. It should be stressed that, although the dispersion of rate parameters can be huge and can comprise many orders of magnitude, the dispersion of structures is small and barely resolvable. The reason is that already very small shifts in intermolecular distances of the order of 0.01\,{\AA} or even less give rise to appreciable energy shifts due to the strong distance dependence of the intermolecular interactions. In addition, rate parameters depend very often in an exponential fashion on energy parameters so that a further strong dispersion arises.\\
Similar to glasses, proteins show spectral diffusion phenomena. Spectral diffusion (SD) stands for the observation that in certain materials the frequency of a spectral line of a chromophore is not constant in time but may undergo spectral jumps. In special experiments spectral jumps up to 100 wavenumbers have been observed.\cite{Hofmann_2003} They can be induced, for instance, by changing the temperature or by light, but they can as well occur spontaneously. The associated time scales cover many orders of magnitudes. Clearly, SD reflects structural randomness, too, because the frequency jumps are the immediate consequence of structural jumps, i.e. of structural changes in a nearby environment of the chromophore. Obviously, metastable structural states do exist. Metastable structural states are a direct consequence of random features in the structure of proteins.\cite{Leeson_1997}\\
An important question in SD-experiments with proteins has always been (and still is) whether the observed phenomena can be directly associated with structural events in the protein and what the role of the solvent and especially the role of the hydration shell is. We will touch on this problem. Apart from experimental observations, structural randomness in proteins is also reflected in molecular mechanics simulations.\cite{Garcia_1997,Becker_1997}
Due to the random features in the energy landscape, proteins share many similarities with polymers and glasses at low temperatures.\cite{Iben_1989} However, despite these similarities, proteins have their characteristic generic features, obviously due to the fact that all of them are characterized by the same hierarchy of interactions: There are covalent interactions, hydrogen bonds, hydrophobic interactions, a series of different electrostatic interactions and all of them have a protein specific interface to the solvent. It is this hierarchy of interactions, fine tuned during evolution, which gives rise to organisation, on the one hand, and to structural frustration\cite{Frauen_1994} and, concomitantly, to randomness on the other hand.\\
We discuss experimental results related to low temperature protein dynamics in terms of two alternative models. One of them (\textit{"random walk on a random path"}) is based on the generalization of the model of protein conformaional dynamics for very low temperatures. The second model (\textit{"independent low energy surface excitations"}) exploits the similarity between proteins and glasses and suggests a modified concept of two level systems in order to explain the observed spectral diffusion. Both models are capable to interpret the anomalous time dpendence of the spectral hole width. The second model is used to interpret additional features including, in particular, structure and temperature dependences. Finally verification experiments are suggested, based on single molecule spectroscopy, which should help choosing the most relevant model out of these two.

\section{SPECTRAL DIFFUSION EXPERIMENTS}
Several types of SD-experiments at low temperatures can be distinguished: Hole burning waiting/aging time experiments\cite{Fried_1986}, hole burning temperature cycling experiments\cite{Koehler_1989} and single molecule waiting time experiments\cite{Hofmann_2003}, as well as various combinations.\cite{Fritsch_1997} The three type of experiments mentioned are frequency domain experiments. In addition, there are also time domain techniques, like stimulated echo experiments.\cite{Leeson_1997} The stimulated echo is the Fourier - transform of the hole burning waiting time experiment with the special feature that very short waiting times can be realized. In the following we will address the frequency experiments only.

\subsection{Hole burning waiting/aging time experiments}
A prerequisite for SD-phenomena to be seen is a structurally unstable environment around the probe molecule. The probe molecule is a dye molecule whose absorption frequency is sensitive to the structure of the environment. In proteins, the probe may be a native prosthetic group or an artificial dye label. Unstable local structures occur as a consequence of structural disorder. The energy landscape acquires many local minima which the system, i.e. the protein, may explore, as time goes on. At low temperature, the exploration of the structural phase space may occur via tunneling processes, but activated processes are not forbidden if the barriers are sufficiently small. Whenever the system changes the structure around the probe, the corresponding absorption frequency undergoes a jump. In ensemble experiments, like hole burning, these frequency jumps give rise to line broadening. The associated width, the so-called SD-width $\sigma$, is the quantity which is measured as a function of time $t_w$. We call $t_w$ a waiting time. It is the time span between burning and reading the hole. The time dependence of $\sigma(t_w)$ reflects the dynamics of the system and may tell us interesting information on the accessible structural states, on the processes involved and on the coupling of the protein to the solvent as well as to the probe. Note that the hole shape which is measured, is a convolution of an initial, time independent, line shape with the waiting time dependent line shape of the so-called SD-kernel. The latter is the interesting quantity and has to be deconvoluted from the measured hole shape. In an hole burning experiment one works close to the homogenous line shape limit, hence, the resolution limit of the experiment is of the order of a few 100\,MHz meaning that structural jumps which lead to spectral jumps in this frequency range can be detected.\\
In addition to the waiting time, we vary a second time parameter, which we call the aging time $t_a$. It is the time span elapsed after the system has reached its final temperature, (e.g. 4\,K), but before it is labeled with a hole. If SD depends on $t_a$, we know for sure that the system is not yet in equilibrium and is still relaxing towards its optimum state. Accordingly, aging and waiting time dynamics reflect relaxation and fluctuation processes. Both types of processes can cover an extremely huge rate dispersion so that low temperature proteins, like glasses, can be safely assigned as non-ergodic.

\subsection{Hole Burning Temperature Cycling Experiments}
This type of experiments has the potential of measuring barriers or even distribution of barriers in the energy landscape of a protein.\cite{Zollfrank_1991} Spectral diffusion is induced by heating the sample. The experiment works in the following way: A hole is burnt at low temperature $T_0$, say 4\,K. Then, the system is subject to a temperature cycle. The cycle is characterized by a temperature parameter which we call the excursion temperature $T_c$. It is the highest temperature in the cycle and is also the parameter which is varied in the experiment. During the temperature excursion the protein may acquire enough thermal energy to cross a barrier to an adjacent energy basin. Upon lowering the temperature, the protein may become trapped in the new basin. When the burning temperature (4\,K) is reached again, the hole has broadened since the protein ensemble attained new conformations which lead to a spread of the initially sharp frequency. In this experiment the hole width is always measured at $T_0$, but as a function of $T_c$. As a rule the experimental times associated with the temperature cycles do not play a significant role as long as we deal with irreversible broadening.
We will present results of these two types of SD-experiments on free base cytochrome - c in various solvents. In addition to the experiments which we performed, we briefly discuss single molecule SD-experiments on light harvesting complexes LH2 of bacterial photosynthesis.\cite{Hofmann_2003}

\subsection{Single Molecule Waiting Time Experiments}
In these experiments a selected spectral line of a single protein is measured by rapidly scanning the line in absorption and detecting the associated fluorescence. The peak frequency of the line is then monitored as a function of a waiting time. Whenever structural events occur in the environment of the chromophore, the frequency undergoes discrete jumps. The final result is a time trajectory of the peak frequency. The jump widths and their probabilities can be determined. As to the experiments with the LH2 complex, a couple of very interesting observations have been made: The jump widths seem to occur in hierarchies: There are very large jumps of the order of the inhomogeneous width. However, they are very rare. Then, there are jumps on an intermediate energy scale and also rather small ones. The probability for smaller jumps is much higher. This hierarchy in the observed jumps was interpreted as reflecting different tiers of the energy landscape. Hence, single molecule SD-experiments were used to map out features of the energy landscape of the protein. In this respect, these experiments resemble the temperature cycle experiments. Whereas in the latter SD is induced by heating the sample, SD is also induced in the single molecule experiments due to the special detection mode. Detecting the fluorescence of a nonresonant transition always implies the simultaneous creation of nonradiative transitions via vibrational relaxation. The corresponding energy release may kick the protein over a sufficiently high barrier so that large frequency jumps can be observed.

\section{EXPERIMENTAL}

\subsection{Sample Preparation}
In all experiments a modified cytochrome - c type protein, mainly free base cytochrome - c (H$_2$-Cc) was used. Iron was removed from horse heart cytochrome - c as previously described.\cite{Vanderko_1975} Trehalose (TH) and glycerol/water (Gl/W) were used as glass - forming matrices. To prepare the protein solution in Gl/W, H$_2$-Cc was first dissolved in potassium phosphate buffer (50\,mM) at pH\,7, and then mixed with glycerol in proportion 1:2.5\,(v/v). Note that in a Gl/W mixture the protein is presumed to be in its native state. To unfold the protein, Guanidiniumhydrochloride (Gua) was used as a denaturant. High concentration (5\,M) of Gua was dissolved in a 2:1\,Gl/W mixture. The protein was then added to the solution. To prepare the deuterated sample, H$_2$-Cc was first dissolved in D$_2$O. To ensure H/D exchange of the amide groups inside the protein core, the protein solution was kept for a couple of days at +4\,C. Protonated glycerol was then added and the resulting mixture was immediately cooled down to 4\,K. Since it takes hours to replace deuterons in the protein pocket, we believe that H/D exchange of the deuterated protein with the protonated solvent may influence the 2 inner hydrogens of the chromophore and the hydrogens on the protein surface, only. Hence, it can be safely assumed that the well protected hydrogens in the protein interior are still deuterated despite the protonated solvent. For all samples the concentration of the protein was around 5-10\,mg/ml.\\
To incorporate the protein in the TH-environment, 5\,mg of H2-Cc were dissolved in 400\,$\mu$l of distilled water and 400\,mg of solid trehalose dehydrate were then added. The solution was aspirated to remove dissolved air and then pipetted on a cover slip. The cover slip was placed on a hot plate kept at about 60\,C for 2\,h. By cooling to room temperature, the "glass" was hard to touch. The optical density in the spectral range, where the hole-burning experiments were performed, was around 0.2-0.3 at liquid helium temperature.

\subsection{Spectroscopy}
Hole burning was carried out with a ring dye laser (Coherent, CW 899-21) pumped at 532\,nm by a frequency doubled neodynium:vanadate laser (Verdy). The width of the laser line was $<$ 1\,MHz. The laser scan range was $\sim$ 30\,GHz. The power levels for burning varied between 5-60\,$\mu$W/cm$^2$. To avoid artificial reburning due to multiple scanning, the reading signal was reduced by more than 3 orders of magnitude. The holes were monitored in the transmission mode and processed by a computer. The resolution time of a hole burning experiment is defined as the time needed to burn and read a hole. Usually, we had to burn deep holes (40$\symbol{37}$) and use averaging techniques to increase the S/N ratio This lowered the resolution time towards the short time cut off to about 5\,min. The widths of burnt-in holes were around 1.5-3\,GHz.
The SD-experiments were performed at 4.2\,K. The time needed for cooling the samples was about 5-20\,min. In the SD-experiments we varied two parameters, namely the aging time ($t_a$) and the waiting time ($t_w$). Several holes within a narrow frequency range of about 50\,cm$^{-1}$ were burnt at different aging times $t_a$. For each experiment the waiting time covered an interval of 2 weeks. By varying the aging time in addition to the waiting time, it is possible to investigate relaxation as well as fluctuation dynamics of the protein at low temperature.
In the thermal cycling experiments the temperature could be controlled within an uncertainty range of 0.1\,K. The temperature sensor was fixed in close vicinity to the sample. Holes were burned and detected at 4.2\,K with burning and reading parameters as mentioned above. We varied only the excursion temperature $T_c$. Due to the limited scan range of the laser we were able to perform temperature cycles up to 40\,K, only. The cycling time $t_c$ was in a 5-15\,min range depending on $T_c$.

\subsection{Data Evaluation}
The quantitative evaluation of a SD-experiment relies on certain assumptions on the SD-kernel. The latter is that portion of the line shape that causes the time evolution of the hole within the experimental time window. The particular line shape of the diffusion kernel depends on the associated dynamics at low temperatures. If SD is the consequence of conformational diffusion of the protein, the diffusion kernel is Gaussian.\cite{Skinner_1999} A model based on a diffusion-like motion is quite in contrast to the so-called standard TLS model that has been worked out to explain low temperature glass dynamics. This latter model is based on quantum jumps in rather localized region, so-called TLS, which interact with the chromophore due to changes in the strain and electric fields. If these fields are of dipolar character, a Lorentzian line shape can be obtained under certain constraining conditions.\cite{Reinecke_1979}  Recent extensive analyses of line shape problems in the framework of non-standard TLS models, predict non-Lorentzian diffusion kernels which, in some special cases may resemble Gaussians.\cite{Kharl_2002} However, it is a difficult task to resolve the line shape problem from an experimental point of view. The main reason is that the SD-kernel must be deconvoluted from the line shape of the burnt-in hole. In all our experiments, the hole measured immediately after burning is usually well fit by a Lorentzian. Note that SD processes may occur in a time window ranging from nanoseconds to practically infinity. Hence, at the first reading, the hole may already contain a contribution due to SD, which may be as large as 50$\symbol{37}$ of the homogeneous line width. For proteins, corresponding values of the order of hundred MHz are typical. This is still much less than the width of the hole. Hence, a Lorentzian line shape can be considered as an appropriate experimental fit function for the initial holes. The experimental time scale of our SD-experiments covered 3 orders of magnitude ranging from minutes to several days. The shape of the hole is a convolution of the spectral diffusion kernel and the initial Lorentzian. To be in line with theoretical reasoning we assumed a Gaussian diffusion kernel. Consequently, in the waiting/aging SD-experiments, the holes were fitted to Voigtians, i.e. to a convolution of an initial Lorentzian with a time dependent Gaussian. Accordingly, the width ( at the inflection point of the Gaussian kernel is plotted as the so-called SD-width. It is necessary to point out that the difference between a Voigtian and a Lorentzian is very small and can experimentally not be distinguished in case the contribution from SD is small. In order to observe such deviations, a high S/N ratio, a rather wide scan range and a good stability of the base line is required.\\
Thermal cycling can potentially help to observe deviations in the line shape due to the much stronger SD in these experiments. As matter of fact, for H$_2$-Cc in the TH-glass and in the deuterated sample non-Lorentzian lineshapes were observed in our experiments. However, it is necessary to stress that thermal cycling experiments may trigger different dynamical processes which may not be observable in waiting time SD-experiments. The change of the line shape due to a temperature excursion has been already discussed.\cite{Kador_1993} The line shape problem is still an open question and deserves special consideration.

\section{RESULTS}
Fig.1 shows the long wavelength regions of the absorption spectra of the same polypeptide, namely H$_2$-Cc under different conditions. Fig.1a represents the spectrum in Gl/W-glass, where the polypeptide is supposed to reside in its native structure. A couple of features should be stressed: The main band shows a clear indication of substructures which we attribute to the tautomeric states of the inner ring protons of the chromophore.\cite{Koehler_1996} The hole burning experiments were performed in the central band in a narrow range around the position marked with an arrow. This band has an inhomogeneous width of roughly 100\,cm$^{-1}$. In addition to the main band there is a broad feature which extends towards lower energies. This broad feature most probably arises from protein aggregates. For the present study, it does not play any role.\\
\begin{figure}
\centerline{\includegraphics[width=3.5in]{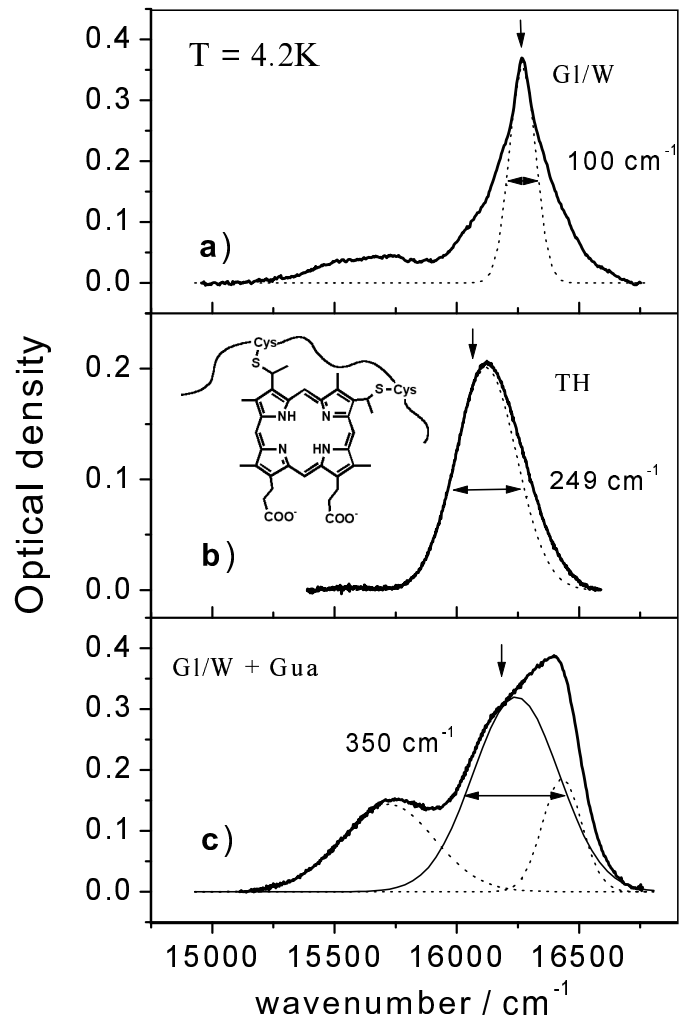}}
\caption{Absorption spectrum of the red-most band of free base cytochrome - c at 4.2\,K under 3 different conditions. The spectra observed in glycerol/water (Gl/W) glass and trehalose (TH) are presented in Figs.1a and 1b, respectively. Fig.1c shows the spectrum in glycerol/water heavily doped with guanidiniumhydrochloride (Gl/W + Gua), so that, in this case, the protein is denatured.
}
\label{fig1}
\end{figure}
Fig.1b shows the spectrum of H2-Cc in a TH-glass. The most obvious difference to the spectrum in Gl/W is the significantly larger inhomogeneous width. There is also a moderate red shift of about 200\,cm$^{-1}$. The shape of the spectrum reflects a rather perfect Gaussian. The small deviations on the high wavenumber side most probably originate from phonons. There is no indication of buried tautomeric states. Again the hole burning experiments were performed in a narrow frequency range around the position marked with the arrow. The insert shows a sketch of the free base chromophore. The polypeptide chain is indicated.
In Fig.1c we show the spectrum of the chemically denatured protein. The solvent is Gl/W with Gua as a denaturant agent. Two features catch the eye: First, the shape of the main band has changed drastically. Clearly, several, most probably slightly shifted tautomeric states build up this band. Again the hole burning experiments were carried out in a narrow frequency range around the arrow. The decomposition of the whole band into single components just serves for estimating the inhomogeneous width of the main component. It is significantly larger than in the native state and comprises some 350\,cm$^{-1}$. Since the fit carries some arbitrariness, this number should be considered just as a guiding number.\\
\begin{figure}
\centerline{\includegraphics[width=3.5in]{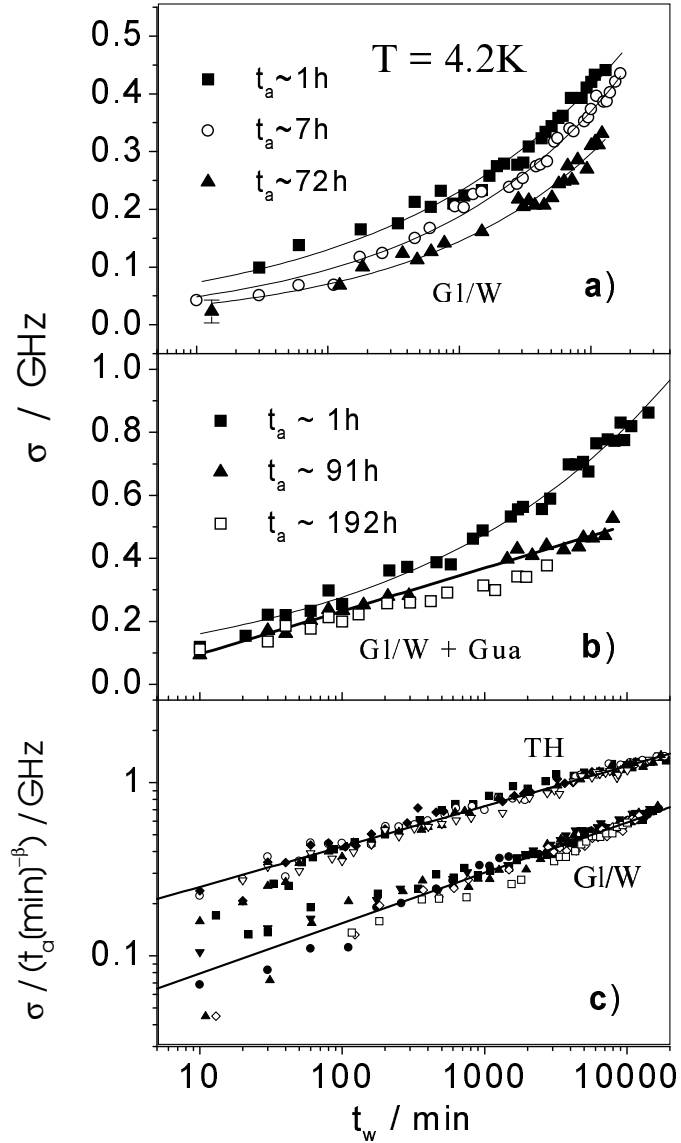}}
\caption{Waiting/aging time spectral diffusion experiments with free base cytochrome - c at 4.2\,K. Fig.2a shows experiments in glycerol/water (Gl/W) glass. Plotted is the width $\sigma$ of the spectral diffusion kernel over the waiting time in the lin-log-representation. Data are shown for three different aging times. These curves are well described by power laws in time. This is shown in Fig.2c in a log-log representation. The data are scaled in a way that the curves for the 3 aging times fall on a master plot. Similar experiments for cytochrome - c in a trehalose (TH) glass are shown for comparison. Fig.2b shows the same type of experiments for the denatured protein in a lin-log representation. For denaturing guanidiniumhydrochloride (Gua) was added. Note that beyond an aging time of about 90\,h, the waiting time dependence of SD changes from a power law behavior to a seemingly logarithmic time dependence.
}
\label{fig2}
\end{figure}
In Fig.2 we represent the SD-waiting/aging time experiments with H2-Cc under different conditions.\\
Fig.2a shows a lin-log plot of the evolution of the hole as a function of waiting time $t_w$ for three different aging times. The solvent is Gl/W. These data unambigously demonstrate that the broadening of the hole is non-logarithmic in $t_w$, i.e. very different from similar experiments in glasses where waiting time dependences $\sim log(t_w)$ have been found.\cite{Breinl_1984,Littau_1991,Schlich_2001} In the protein, they rather follow a power law with an exponent of 0.29. The data demonstrate as well that aging phenomena take place: The SD-width decreases as the time $t_a$ between cooling and hole burning increases. We conclude that the degrees of freedom responsible for SD have not yet reached a stationary behavior.
For comparison, the behavior of the chemically denatured protein is shown in Fig.2b, also for three largely different aging times. We find that the SD-width for the shortest aging time (1\,h) is significantly larger than in the native state, namely by almost a factor of 2. The time evolution, however, is still governed by a power law with the same exponent as in the native state. The most remarkable observation, however, concerns the dramatic change of the time evolution of the hole width beyond aging times of about 100\,h: The power law changes to a seemingly logarithmic time behavior (at least in the experimental time window) as is immediately obvious from the straight line representation in the lin-log plot.\cite{VVP_2004} Note that since a power law is exponential in $\ln(t)$, the change is really dramatic.\\
In Fig.2c we compare the data for Gl/W with data for TH. In both cases the protein is in its native state. The representation is on a log-log scale to demonstrate the power law behavior of the SD-width in this protein. The data are scaled in a way that the different sets obtained for various aging times (e.g. Fig.2a) fall onto a master plot.\cite{Schlich_2000} The exponents for the two solvents (Gl/W and TH) are slightly different, 0.29 and 0.23, respectively.\cite{VVP_2002} Notice that there is a remarkable solvent effect in the magnitude of SD: The SD-width in TH is larger by almost a factor of 2.\\
\begin{figure}
\centerline{\includegraphics[width=4in]{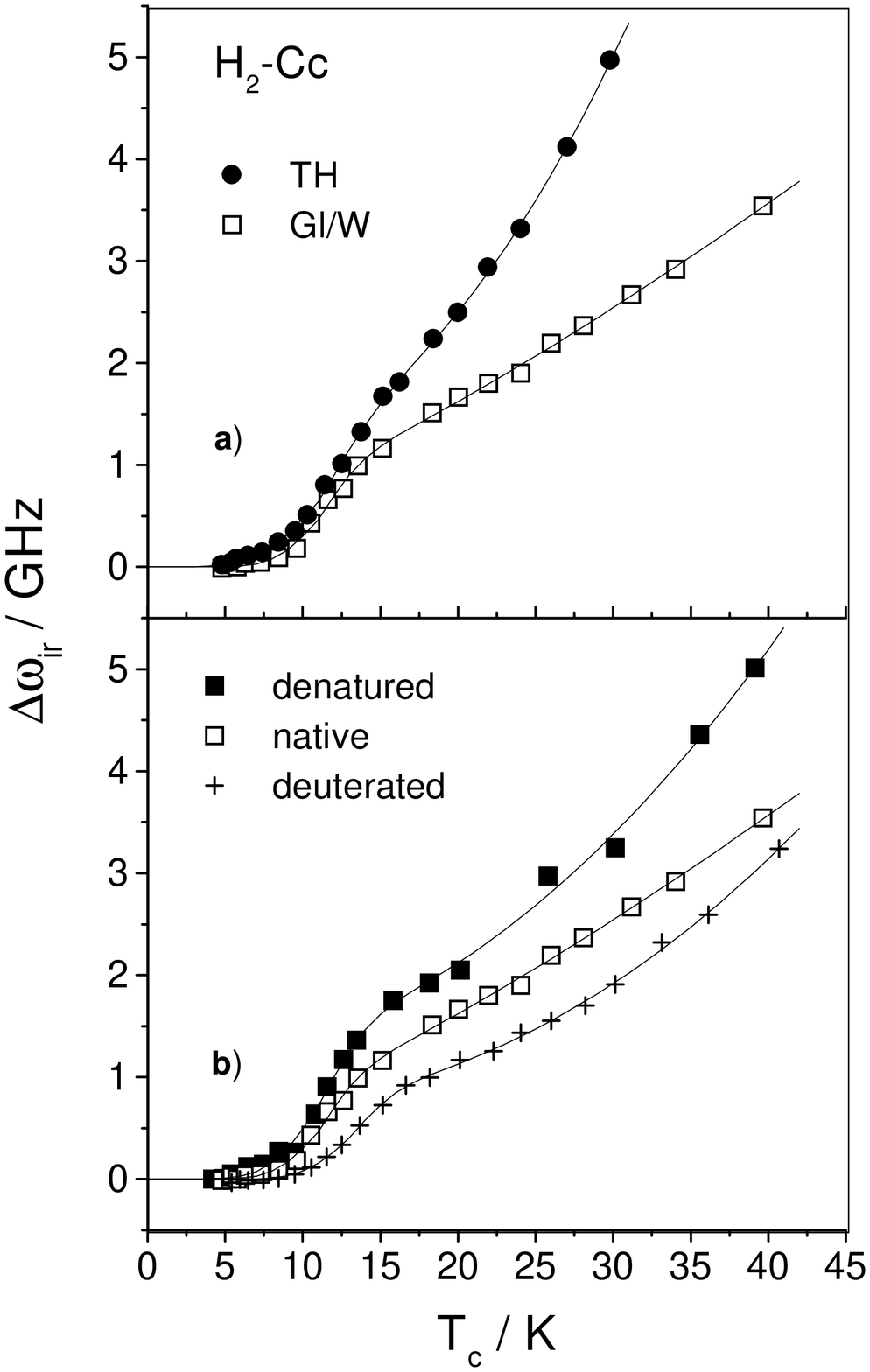}}
\caption{Thermal cycling SD-experiments. The broadening of the hole $\Delta\omega_{ir}$ is plotted over the excursion temperature $T_c$, the highest temperature in a thermal cycle. Points are experimental data, solid lines the result of fitting as explained in the text. The thermal cycling SD-behavior in trehalose (TH) and in glycerol/water (Gl/W) is compared in Fig.3a. Fig.3b allows the comparison of the thermal cycling SD-behaviors observed for the denatured protein, the native protein in Gl/W and for the deuterated protein in Gl/W. Note that thermally induced irreversible broadening exhibits a step-like behavior as $T_c$ increases.
}
\label{fig3}
\end{figure}
Data from thermal cycling SD-experiments are shown in Figs.3a and b. In Fig.3a we compare the behavior in the two solvents Gl/W and TH. Again, we observe a remarkable solvent effect: The thermal cycling SD-width in TH is significantly larger than in Gl/W, however, only above an excursion temperature of about $T_{co}\approx$ 12\,K. Below this value the differences are marginal. We stress two interesting observations: First, above $T_{co}$ the dependence of SD-broadening on excursion temperature is superlinear with exponents between 1.3 and 2.3, and, second, the thermal cycling SD-width shows at least one step which occurs around 12\,K. As to the step, we note that there is a slight dispersion of the step temperature $T_{co}$ (11 K for Gl/W, TH and the denatured protein, 13\,K for the deuterated protein) as well as a slight dispersion of the respective widths (2-3\,K).\\
There is also a denaturing effect as well as a deuteration effect: SD is larger in the denatured sample and smaller in the deuterated sample as compared to the Gl/W case.\\
In the following discussion we want to develop model guide lines along which this rich scenario of SD-effects might be understood. We stress, however, that the body of data is still not large enough to arrive at unambiguous conclusions.

\section{DISCUSSION}

\subsection{The Protein State of Matter at Low Temperature: Characteristic Spectroscopic Features.}
SD-phenomena have been discovered in glasses long before they were shown to occur also in proteins.\cite{Breinl_1984} The first experiments in proteins were of the temperature cycling type. They were performed with antenna pigments of bacterial photosynthesis.\cite{Koehler_1989_a} Spectral diffusion was clearly there, but the temperature behavior was very similar to that observed for glasses. Accordingly, the results were interpreted using models developed for the glassy state, namely the standard TLS-model disregarding the fact that proteins are finite systems and non-homogeneous. However, the existence of TLS-states in proteins has already been postulated in earlier experiments on their specific heat as well as on their dielectric properties.\cite{Singh_1984,Yang_1986} From the early spectral diffusion experiments, it was not quite clear what the influence of the glass matrix is. In the antenna proteins investigated the chromophores are near the protein surface, hence, may feel the influence of the solvent to a large extent. To overcome this problem, experiments were conducted with modified heme proteins, such as free base cytochrome - c, free base myoglobin or free base horseradish peroxidase as well as with some metal derivatives of these proteins.\cite{Schlich_2001_a,Schlich_2001_b} In these proteins the chromophore is buried in the so-called heme pocket, hence, could be expected to be decoupled from the solvent by the protein environment to a much higher degree. Indeed, quite specific features were observed, distinctly different from glasses. In thermal cycling experiments kind of discrete steps could be identified, and in waiting time experiments time laws drastically different from glasses were observed. The spectral diffusion dynamics in glasses follows logarithmic time laws whereas in proteins power laws with a rather low exponent of order 0.25 were observed.\cite{Schlich_2003} In addition, SD was found to be sensitive to deuteration\cite{Schlich_2001_b} and, despite the fact that the chromophore was embedded in the interior of the protein, solvent effects did occur.\cite{VVP_2002,Schlich_2001_c,Schlich_2002} The question then was how the special features measured could be explained and what kind of protein properties could they reflect. The first approach was to view spectral diffusion as the direct manifestation of conformational diffusion in a complex structural phase space\cite{Berlin_1996,Berlin_2001}. We called this approach the \textit{"conformational diffusion model"}. The diffusion model was completely devoid of any relation to TLS. Under certain restrictive assumptions, a power law with the appropriate exponent could be explained (see below).\cite{Skinner_1999, Burin_2002} There are, however, two problems: One problem concerns the line shape, the other problem concerns the solvent effects. In the conformational diffusion model solvent effects can be explained only in a rather unspecific way in terms of the glass transition temperature and its influence on the hole width.\cite{VVP_2002} In addition, an influence of the solvent on the frequency correlation time had to be postulated.\cite{Schlich_2003,Schlich_2001_c} As a second approach, we present here a model which is based on the presence of TLS, but which takes into account specific features of globular proteins, namely their finite, rather well defined size and the presence of rather spherical interfaces between protein and solvent. We call this approach the \textit{"independent low energy surface excitation"} model. A major influence of surface TLS on SD was already considered in earlier paper on SD phenomena in proteins.\cite{Hartog_1999}

\subsection{Randomness in Proteins as Compared to Glasses}
Despite their differences, low temperature proteins and glasses share essential characteristics. For instance, both show non-exponential relaxation features and non-Arrhenius behavior.\cite{Iben_1989,Berlin_1995} Both features originate from a distribution of barriers. Proteins show also a so-called dynamic transition which has many similarities with glasses. At the dynamic transition large amplitude motions with quasi-diffusive character become frozen in.\cite{Vitkup_2000,Doster_1989} Proteins as well as glasses are described by energy landscapes with random features. At sufficiently low temperatures, both systems are non-ergodic. As to the differences, we stress the finite size of proteins and their nonhomogeneous character, the fact that there are quite specific interfaces to the solvent and the hierarchy of interactions. The transition into the native state is not a glass transition either, but is of the 1st order. These differences may be responsible for the characteristic differences in SD-dynamics.\\
Most features of SD in glasses are well explained within the frame of the standard TLS-model.\cite{Fritsch_1996} Almost all phenomena observed can be reduced to 3 properties of the standard model: The almost constant density of TLS-states (note the pseudogap\cite{Salvino_1994, Burin_1995,Burin_1998} formed at low energies through TLS interaction, which is significant for our interpretation of aging, see below), their dipolar coupling to the probe and their homogeneous spatial dispersion in a quasi infinite space. These properties lead to a Lorentzian line shape of the SD-kernel and to a logarithmic time dependence of SD-broadening. We have performed many comparative experiments to unravel specific features of proteins as compared to amorphous organic materials. For instance, we performed optical Stark-effect experiment to show that proteins have symmetry breaking properties\cite{Koehler_1996} due to their well organized structure. Recently we performed comparative SD-experiments for protonated and deuterated Gl/W-glass, doped with a dye probe and compared the corresponding behavior with the SD-pattern in a protein. The results clearly showed that proteins do have a specific dynamics. In addition, these experiments also showed that a direct coupling of the TLS of the glass matrix to the chromophore in the heme pocket is marginal.\cite{Schlich_2001}

\subsection{Model considerations}

\subsubsection{A) Conformational Diffusion: Random Walk on a Random Path}
The conclusion from above observation is that the standard TLS model\cite{AHW_P_1972} is not capable of describing low-temperature SD in folded proteins. In fact, the time dependent broadening of a spectral hole shows an anomalous power law behaviour (Fig.2)
\begin{equation}
\sigma(t)\sim t_w^{\alpha} 
\end{equation}                                                                                  
with the universal exponent $\alpha$ in the range
\begin{equation}
0.2<\alpha<0.3 
\end{equation}
observed in a waiting time window from minutes to weeks. This behavior contradicts the standard TLS model which predicts either linear or logarithmic time dependences.\cite{BH_1977,HR_1986} The power law behavior in equ.~(1) has the form of an anomalous diffusion with a remarkable memory. In other words, the fluctuations can be associated with conformational motions characterized by a large probability of returning back to the initial state. Such a behavior was realized in models of conformational diffusion of proteins at relatively high temperatures.\cite{Berlin_1995,Berlin_1997} In these models protein conformational states are represented as energy minima of various depths and the conformational dynamics results from thermally activated transitions between these minima. Protein dynamics can then be described by an anomalous diffusion law, equ.~(1), with the exponent $a$ directly proportional to the temperature $\alpha = T/T_g$, where $k_{B}T_g$ is the characteristic potential barrier separating two adjacent conformations. One can expect that at very low temperatures $T \sim $ 1\,K, a thermally activated motion is replaced by tunneling and the exponent $\alpha$ becomes temperature independent. It is not clear, however, why this exponent should have an universal value of about $0.25$ giving rise to the spectral diffusion behavior as described in equ.~(1).
One possible scenario leading to the exponent $\alpha\approx{0.25}$ has been suggested.\cite{Burin_2002} This scenario uses the arithmetic identity $\alpha = 0.25 = 0.5/2$, where the exponent 0.5 characterizes the classical random walk behavior. Consider the time dependent displacement of a particle performing a random walk on a random path (Fig.4) from its initial position $\bf0$. Let the average time of a single step be $\tau_0$. During the time $t$ the particle makes $t/\tau_0$ steps in right and left directions. Each time the direction along the path is chosen randomly, so the mean square displacement of the particle along the random path is given by the random walk expression $N_t\sim \sqrt{t/\tau_0}$. The displacement of the particle in real space is given by the sum of $N_t$ random vectors along the random path (these vectors are shown by arrows in Fig.4). Since the sum of these $N_t$ random vectors represents the random walk, the displacement from the initial position $\bf 0$ scales as the square root of the number of steps which is $\sqrt{N_t}\sim t^{\;0.25}$. Below we discuss the implication of this picture on SD in proteins at low temperature.\\
\begin{figure}
\centerline{\includegraphics[width=4in]{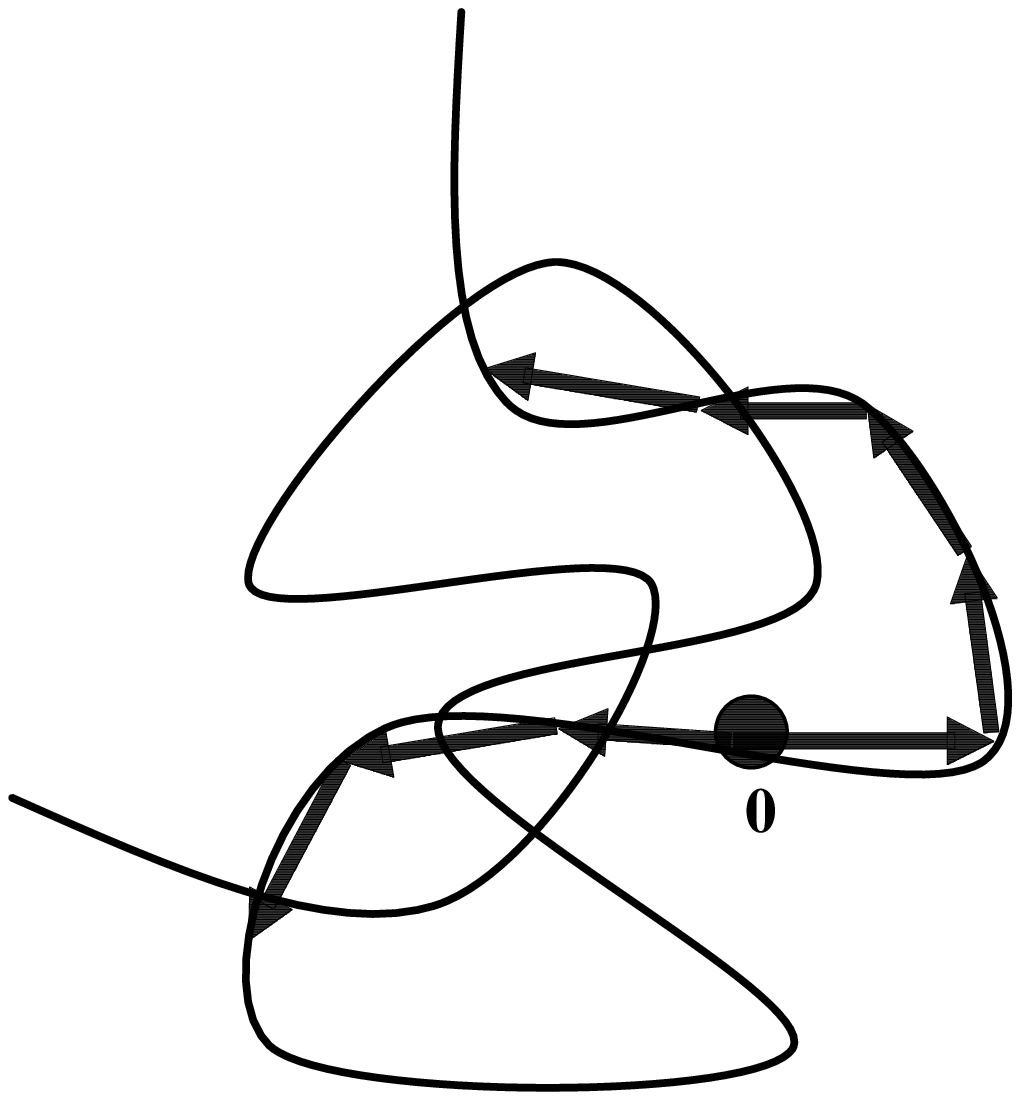}}
\caption{Random walk on random path. The particle starts its motion at the point designated by {\bf 0} and the dark circle and makes the number of discrete steps indicated by arrows in right and left directions.
}
\label{fig4}
\end{figure}
Consider further the conformational diffusion of the protein and its influence on the absorption energy of the chromophore. Let the protein make tunneling jumps between adjacent conformational states of close energies during the time $\tau_0$. Each jump causes a random change of the absorption energy by a characteristic value $U_0$ in arbitrary direction. The average value of the energy change is zero, while its average squared value is defined as $U_0^{\;2}$. Assume further that the absolute value of $U_0$ does not strongly fluctuate for different transitions, so that we can describe the overall absorption energy change, caused by many transitions, in terms of the law of large numbers. Note that this assumption is not anymore true if the conformational transitions are local and random in space (like two level systems) and if they affect the electronic energy through dipolar or elastic interactions. In that case the most significant change in the electronic energy comes from the transition which is closest in space.\cite{BH_1977} To avoid this situation and to justify the law of large numbers, we can either assume that each conformational transition leads to the rearrangement of the whole protein molecule or, alternatively, that the relevant transitions occur at the interface between protein and host glass. This assumption is similar to the concept of non - universal surface TLS as suggested by Neu and Heuer based on results of molecular dynamics simulations.\cite{Neu_1997} Then the law of large numbers will produce a Gaussian shape of the SD-kernel which was also used in the analysis of the experiments.\cite{Skinner_1999}
To justify the picture of a \textit{"random walk on a random path"} one should take the effective dimensionality of the protein conformational space to be close to one. This assumption can be justified by percolation theory arguments. At high temperature the conformational space has a very large effective dimensionality. As the temperature goes down, the range of accessible energies is remarkably reduced, and, consequently, the number of accessible conformational states is reduced. In the low temperature limit the conformational transition from a given state is possible to only a few neighboring conformations within an energy range of no more than the thermal energy. In the liquid He temperature range these transitions cannot be thermally activated, and, consequently, should be associated with quantum tunneling. The tunneling rate is exponentially sensitive to the energy difference between conformations. Therefore, only transitions to a minimum number of neighboring conformations, namely to those which are energetically acceptable and which lead to irreversible relaxation, should be considered. These conformational states must form the infinite percolation cluster in order to make the relaxation irreversible. The dimensionality of this cluster is close to unity even for a high dimensional conformation space.
With the above assumptions one can approximate SD by a random walk on a random path (Fig.4). The random walk occurs along the quasi-one-dimensional infinite percolation cluster of accessible states in the low temperature conformational space. In this case the effective number of steps out of the initial point scales with the time as $N_t\sim \sqrt{t/\tau_0}$. The random coil geometry is associated with a random change of the absorption energy of the chromophore by approximately $U_0$ for each tunneling step. Then this mechanism will lead to the characteristic change in energy $\sigma(t)\sim U_0\cdot\sqrt{N_t}\propto t^{\;0.25}$ which fits well to the experimental results.\\
An experimental verification of the above scenario is a hard problem. In fact the conformational space model does not tell us much about the nature of the conformational transitions or about their interactions with the chromophore. However, some mathematical implications of the model could, at least in principle, be tested in hole burning as well as in single molecule experiments. For instance, the distribution of energy fluctuations induced by a \textit{"random walk on a random path"} is not a single Gaussian. Instead it is a superposition of Gaussians characterized by a width dispersion. The width of this dispersion is defined by the range of the numbers of successful steps for each individual protein. This range extends from zero up to the characteristic value $\sigma \sim t^{\;0.25}$.\cite{Patash} Experimentally, however, it is very challenging to analyze the shape of a hole in all these details. Hence, an experimental verification of the random coil model is still open.\\
In single molecule experiments of proteins, it is the statistics of the energy jumps that can be investigated and used to test the model. The small energy changes should dominate in the overall energy fluctuation compared with the rare large jumps. In addition, in a quasi-one-dimensional conformational space a memory effect should show up. This means that the absorption energy should periodically return back to its exact initial value.\\
Irrespective of whether the model can be checked experimentally or not it does not tell us anything about the microscopic nature of the anomalous spectral hole broadening in proteins. For this purpose a quantum mechanical model is necessary in order to understand and interpret the whole set of effects including solvent effects, protein denaturing effects, deuteration effects, temperature dependencies, etc., which are experimentally observed. Below we discuss a microscopic scenario based on possible modifications of the model of standard two level systems\cite{AHW_P_1972} which can be derived from the anomalous aging behavior. This scenario looks like a solid alternative to the \textit{"random walk on a random path"} model. However, under certain conditions both models can show similar behaviors.

\subsubsection{B) Surface TLS Bring About Quite Specific Features: Independent low energy surface excitations}
The aging effect of spectral holes observed in native proteins\cite{VVP_2004,VVP_2002} is anomalously weak compared to the expectations of the standard TLS model. In the standard TLS model, spectral hole broadening is associated with the dynamics of TLS having energies comparable with the thermal energy and higher. The contribution of thermal TLS is fluctuational because they can occupy both, excited and ground states. The contribution of higher energy TLS is relaxational. Those of them having relaxation times longer than the aging time, can be in their excited states at the beginning of the spectral hole burning experiment. They contribute to the spectral diffusion only once, namely when they irreversibly approach their ground states. The phase volume of this subgroup of TLS, with energies exceeding the thermal energy, is larger than the phase volume of thermal TLS which are responsible for the equilibrium fluctuation effect in SD. Therefore the aging effect should be comparable or even larger than the contribution from the fluctuations.\\
The aging effect in SD for native proteins can be expressed by a power law in the aging time ta
\begin{equation}
\sigma \sim t_a^{-\beta}, \beta \approx 0.07
\end{equation}
It is very weak compared to the expectations of the standard TLS model. However, equ.~(3) looks quite similar to the logarithmic relaxation of the low energy density TLS due to the dipole pseudogap which was discovered earlier in non-equilibrium dielectric measurements in glasses.\cite{Salvino_1994,Burin_1995,Burin_1998}\\
This similarity brings about the suggestion that the aging effect in SD of native proteins may be caused by the low energy excitation dynamics rather than by the transitions acquiring or releasing thermal or even higher energy as in the standard TLS model.\cite{HR_1986} In other words we assume that the transitions contributing to the spectral hole width involve changes in energy smaller than the thermal energy because the transitions with higher energy are too slow. We argue that the aging effect in SD is proportional to the density $g_o$ of these low energy excitations. These excitations are coupled to each other as well as to the higher energy excitations by elastic and/or electric dipolar interactions. At low temperatures, as in our experiments, the interactions reduce the low energy density of these states as was shown experimentally\cite{Salvino_1994,Burin_1998} and described theoretically\cite{Burin_1995,Burin_1998} based on the so-called dipole gap model. It was shown that the density of excitations under non-equilibrium conditions, $g_o(t)$, formed by a suddenly applied external DC field, decreases logarithmically with time due to relaxation (aging) processes as
\begin{equation}
g_o(t)=g_o(0)\cdot (1-\chi\cdot \ln(t/\tau_0))
\end{equation}
Here $\tau_0$ is a characteristic minimum relaxation time of the excitations involved and $\chi$ is a small parameter, characterizing the interaction between different excitations. The absolute value of $\chi$ is of order 0.01, and it changes from system to system. This parameter can as well be in a range around 0.07, the value necessary for matching the aging behavior as described in equ.~(3).\\
In the dipole gap model, fast transitions take place, if the resulting energy change is small. How can this be realized in two-level excitations? Clearly, this is not the case for transitions involving single phonon absorption or emission. For these transitions the rate is directly proportional to the energy change.\cite{HR_1986} Therefore they are most efficient if the energy splitting is greater or equal the thermal energy.\\
However, the situation is quite different if multi-phonon transitions are involved. For very small tunneling amplitudes, these transitions can make the main contribution in the resonant tunneling processes between nearly degenerate states. For instance, they govern the quantum diffusion of $^3$He impurities in $^4$He-crystals with the characteristic tunneling amplitude $\Delta_0 \sim k_B \cdot 10^{-4}K$,\cite{Kagan} or the diffusive motion of ortho-hydrogen in a para-hydrogen lattice,\cite{Kokshenev} and, possibly, high spin tunneling in interacting Mn$_{12}$Ac magnetic molecules.\cite{Tupitsyn} In the two latter cases the $\Delta_0$ values are of the order of $k_B \cdot 10^{-9}K$.\\
To understand why multiphonon transitions prevail at low temperature, we consider the transition rate of a TLS with asymmetry $\Delta$ and tunneling amplitude $\Delta_0$, as in the standard TLS model. Following Kagan and Maksimov\cite{Kagan} (for biophysical applications of their results see also\cite{Berlin_1996_a}), one can decompose this rate into two contributions of incoherent (single phonon) and coherent (multi-phonon) channels
\begin{equation}
\kappa_{tot} \approx \frac{A}{\hbar}\sqrt{\Delta^2+\Delta_0^2} \coth \left(\frac{\sqrt{\Delta^2+\Delta_0^2}}{k_BT} \right)\Delta_0^2 + \frac{1}{\hbar} \frac{\gamma\:\Delta_0^2}{\gamma^2+\Delta_0^2+\Delta^2} 
\end{equation}
The first term here represents the single phonon transition accompanied by emission or absorption of one phonon. The second term stands for the resonant transition rate, realized for excitations with very small asymmetry $\Delta$. The factor $\gamma/\hbar$ is the dephasing rate of TLS induced by multiphonon scattering without transitions.\cite{Tupitsyn} Note that the second term in equ.~(5) has the standard form of a resonant scattering rate.\\
For almost degenerate excitations, i.e. if the asymmetry goes to zero, the characteristic transition rate becomes very large and reaches the scale of $\Delta_0^2/\gamma\hbar$, which is much greater than the thermal TLS relaxation rate $A\Delta_0^2k_BT/\hbar$ in the low temperature limit because of the fast decrease of the dephasing parameter $\gamma/\hbar$ with temperature.\cite{Kagan} We assume that, within the time window of our experiment, the tunneling amplitude $\Delta_0$ is so small that the single phonon processes involving thermal or even higher energy changes, can be fully neglected. Note, that a similar condition holds for quantum diffusion of ortho-hydrogen in a para-hydrogen matrix, where typical time constants for single phonon processes are of the order of tens of years, whereas the time scale for resonant multi-phonon transitions are of the order of hours.\cite{Kokshenev,Li}\\
We believe that a similar situation could be realized in proteins.\\
If the frequency fluctuations in the absorption spectrum of the chromophore are associated with the low-energy transitions, then the number of transitions is directly proportional to the density of low energy excitations. In the non-equilibrium regime this density is affected by the dipole gap relaxation (see equ.~4). Note that equ.~(4) is formally a first order expansion of the power law behaviour (equ.~3), and hence this equation describes the observed aging behavior. Now the question arises of whether one can also successfully interpret the anomalous spectral hole broadening (equ.~1) with waiting time in terms of the same model?\\
The answer to this question is positive. In the following we discuss how the SD-behavior can be interpreted within a model based on low energy excitations. A possible microscopic understanding of the nature of these low energy excitations can be attained by making use of the concept of surface TLS.\cite{Neu_1997}
First, we demonstrate that, this simple toy model yields the $t^{\;0.25}$-law for the SD-rate in a natural way.
Assume that the chromophore interacts with special two level transitions which are all characterized by the same tunneling amplitude $\Delta_{0*}$. This assumption is the opposite limit as compared to the logarithmically broad distribution of tunneling amplitudes in amorphous solids.\cite{AHW_P_1972}
Assume further that $\Delta_{0*}$ is sufficiently small so that single phonon processes are negligible during the whole experimental time, similarly to Refs.~[54,55]. Then, the transition rate is given by the second term in equ.~(5)
\begin{equation}
\kappa_{tot} \approx \frac{1}{\hbar} \frac{\gamma\:\Delta_{0*}^2}{\gamma^2+\Delta^2} 
\end{equation}
During a waiting time $t_w > \hbar\gamma/\Delta_{0*}^2$ those excitations can undergo transitions whose relaxation rate is larger than $t_w^{-1}$. These excitations satisfy the condition
\begin{equation}
\frac{\gamma\:\Delta_{0*}^2}{\hbar\Delta^2}t_w>1
\end{equation}
The inequality (7) restricts the appropriate asymmetries to a range
\begin{equation}
\left|\Delta\right|<\sqrt{\frac{\gamma\Delta_{0*}^2t_w}{\hbar}}
\end{equation}
The total density of excitations which undergo transitions during a time $t_w$ is proportional to the size of the domain of allowed asymmetries (equ.~8)$(-(\gamma\Delta_{0*}t_w/\hbar)^{1/2},(\gamma\Delta_{0*}t_w/\hbar)^{1/2})$ , and, consequently, we have
\begin{equation}
N_t \sim g_0 \sqrt{\frac{\gamma\Delta_{0*}^2t_w}{\hbar}}
\end{equation}
where $g_0$ is the density of TLS states described by equ.~(4). Note that it contains the contribution due to aging.
Similarly to the model of a \textit{"random walk on a random path"}, we assume that each low energy transition randomly disturbs the absorption energy of the chromophore by some characteristic energy $U_0$. Since this energy change occurs in random directions, the overall spread of the absorption line induced by $N_t$ transitions can be estimated as
\begin{equation}
\sigma(t) \sim U_0\sqrt{N_t} \sim U_0\  g_0^{1/2}\left(\frac{\gamma\Delta_{0*}^2}{\hbar}\right)^{1/4}\cdot t^{\;0.25}
\end{equation}
Accordingly, within the framework of our model, the SD-width of the hole scales with $t^{\;0.25}$ in agreement with the experiments. The aging effect is incorporated in Eq.~(10) through the time dependent density of states $g_0(t_a+t_w)$, as given by Eq.~(4). This leads to the factor $\sqrt{1-\chi\ln((t_a+t_w)/\tau_0)}$ in equ.~(10) accounting for the weak dependence on aging time (equ.~3) as was found experimentally.
Similar arguments allow comments on some experiments aimed to investigate the broadening of a spectral hole induced by an external DC-field in various time ranges.\cite{Wunder_1998} The field produced a much stronger effect than measured in the protein aging behavior (equ.~3). We attribute this external field effect to the portion of the excitations that attain a low energy when the field is switched on. At zero field these excitations are characterized by a relatively large asymmetry $\Delta$ (cf. equ.~6) so that multi-phonon transitions are prohibited. However, they change their states during the application of the field so that multi-phonon transitions may become very effective. If the field is turned off, they become immobilized again and freeze in a non-equilibrium configuration. Accordingly, the fluctuations of these special excitations produce the irreversible and possibly large contribution to the hole width that is absent in aging phenomena of proteins.\\
In the following we briefly address the line shape problem associated with our special model on low energy excitations: If many excitations contribute to the line broadening in a uniform fashion, the resulting change in the absorption energies obeys the law of large numbers. Therefore, one expects a Gaussian shape for the SD-kernel of the spectral hole, even for dipolar interactions between TLS and chromophore. The only restrictive condition is that the probability of finding a TLS in very close vicinity of the chromophore is low.\cite{Kador_1993}
So far we did not address the question of the microscopic nature of the special TLS in our model which are responsible for low temperature protein dynamics and which are characterized by almost identical tunneling amplitudes. One possibility is that they are associated with tunneling transitions between different conformations of the protein. Checking out whether this is true or not requires special investigations. Another possibility comes from the concept of surface TLS, which was introduced by Neu and Heuer [51] based on their molecular dynamics investigation of a large chromophore in a glassy host. They demonstrated that a new class of two level excitations is formed at the interface between glass and chromophore which are very different from the standard TLS.\cite{AHW_P_1972} They have the necessary narrow distribution of tunneling amplitudes with a magnitude smaller than the maximum tunneling amplitudes of the standard TLS. At T $\sim$ 4\,K, the latter quantity is of order $\Delta_0\sim k_B \cdot$ 1\,K.
The surface TLS are located at the protein-glass interface. Hence, for globular proteins with the chromophore roughly in the protein center, their interactions with the chromophore are all of the same order of magnitude. Thus, one can apply the law of large numbers to estimate their overall effect. In addition, since their distribution of tunneling amplitudes is narrow, we can use our model with all identical tunneling amplitudes $\Delta_{0*}$, at least for approximate evaluations. As a consequence, surface TLS seem to be good candidates for the special low energy excitations in our model, and we can use their multi-phonon dynamics to interpret the anomalous time dependence of the spectral hole width (equ.~1).\\
One should note that other inhomogeneous glasses including for instance highly porous vycor glasses\cite{Nittke,Nittke_a} and small glassy particle arrays\cite{Watson} also show remarkable deviations from the standard TLS model. Perhaps, their behavior can be explained by surface TLS as well. Whether the anomalous dynamics typical for proteins, is observed for these system too has to be checked. However, it can not be excluded that some of the specific features of the model, especially the large number of TLS, are intimately connected with the special nature of protein surfaces.\\
It should be emphasized that the model of independent excitations, equs.~(7, 8, 9, 10), is not the only possible way to interpret SD-experiments using multi-phonon resonant transitions. An alternative possibility is based on spectral self-diffusion within the interacting ensemble of surface excitations: Each transition may change the energy of other excitations. This process may bring some excitations to the low energy resonance domain where multi-phonon transitions (equ.~7) are really fast, while other excitations can be thrown away from their resonances. A multiple repetition of this process gives rise to a self-consistent spectral diffusion (equ.~1) with the reasonable value of the exponent $a$, close to the experimental observations.\cite{Prokofev} Contrary to the model of independent excitations, a remarkable memory effect in SD should be seen similarly to the \textit{"random walk on a random path"} model.\\
Single molecule spectroscopy can be used to test the above model. An important difference as compared to the model of \textit{"random walk on a random path"} is the absence of memory effects. In fact there are no correlations in excitation transitions. As a consequence, the probability for the protein molecule to return back to exactly the same configuration characterized by the same absorption energy should decrease with time very fast.
For further tests of the model, it is always important to check the dependence of the observable parameters (e.g. the spectral hole width) on as many external conditions as possible, for instance, on the temperature, on the solvent, on H/D-exchange, on the protein state (e.g. native vs. denatured), etc. The model of low energy excitations permits such an analysis on a qualitative level. For instance, from equ.~(10) one can predict the temperature dependence of the SD-width
\begin{equation}
\sigma(t)\sim \gamma ^{\;\frac{1}{4}}
\end{equation}
The temperature dependence of the multi-phonon dephasing rate $\gamma$ is given by a power law $\gamma \sim T^{\:n}$ with exponents $n~=~5$,\cite{Kokshenev} or $n~=~7$,\cite{Nittke} depending on the symmetry of the problem. In both cases the temperature dependence of the SD-width is given by a power law $\sigma \sim T^{\:\eta}$ with an exponent $\eta$ between 1 and 2. This does not differ strongly from the linear temperature dependence, which is usually used to interpret experimental data.\cite{Haarer}\\
Another important response to changing external conditions is the increase in inhomogeneous broadening as well as in SD if the solvent is changed from Gl/W to TH. The main difference between these two systems is the presence of water in the protein pocket\cite{Scharnagl} in Gl/W, while the TH solution is entirely dry. The presence of water molecules in the pocket, should increase the dielectric constant in Gl/W as compared to TH due to polarization of the water molecules. Note that there is experimental evidence that in water containing solvents, water molecules inside the protein are still mobile down to temperatures in the Kelvin range.\cite{Vanderko} Assume that SD arises from dipole-dipole interaction of surface TLS with the chromophore, located inside the protein. We further assume that the elastic interaction is less significant for surface excitations because the coupling between the atoms of the protein with the atoms at the interface is weaker than the bulk coupling. Then, the dominating electrostatic dipolar coupling may be reduced in Gl/W because of an increased dielectric screening as compared to TH. As a consequence, a larger SD and possibly a larger inhomogeneous broadening may result in a TH-solvent.\\
As to the strange behavior of the denaturated protein which changes its time dependence of SD-broadening from a power law behavior to an almost logarithmic behavior for sufficiently long aging times, we note that the model of independent uniform low energy surface excitations may break down. A denatured protein is supposed to have a random coil-like structure. Hence, its surface is not as well defined as in globular proteins. The coupling of the surface TLS to the chromophore is no longer uniform because there may be a distribution of distances. Moreover, the chromophore itself becomes exposed to the solvent and may feel the bulk excitations. However, the question why the behavior changes with aging time is still open.

\subsection{Comments on Temperature Cycling SD}
Temperature cycling SD-experiments have the capability of mapping out features of the energy landscape of the protein. The interesting observation concerns the steps in all the experiments shown in Fig.3. The step around 12\,K is rather pronounced. The data show that below the "step temperature" thermal cycling is almost reversible, i. e. the acquisition of new configurations which induce line broadening, is much less pronounced than above the "step temperature". Above the step temperature thermally induced SD is strong and shows a pronounced solvent effect: In TH the broadening is much stronger than in Gl/W. From the fact that we have a significant solvent effect above 12\,K but only a marginal solvent effect below 12\,K and from the observation that the step occurs in both solvents at the same temperature, we conclude that the degree of freedom which is triggered at the step temperature involves a protein associated barrier which is not influenced significantly by the solvent. We stress that in both solvents the protein is supposed to be intact.\\
In Fig.3b we compare the thermally induced SD pattern of the chemically denatured state with that obtained for the deuterated sample. For comparison the SD-behavior in Gl/W is shown as well. An eye-catching feature is the fact that in the chemically denatured state, the step occurs at the same temperature as in the native state. This observation leads immediately to the conclusion that the step corresponds with a structural barrier which is obviously not destroyed in the denaturing process.\\
From this observation we can draw a couple of interesting conclusions. First, the energy landscape of the protein below an energy scale as given by $k_B \cdot $12\,K is practically smooth on the energy scale of the the reading temperature $T_0$. This means that barriers are smaller than $k_BT_0$. This conclusion follows immediately form the observed reversibility and is also in line with earlier experiments.\cite{Fritsch_1997} Second, above the step temperature, the landscape is rough, i.e. there are barriers larger than $\sim k_BT_0$. For a rough estimation, we can relate the spread of the energy $\Delta\epsilon$ of the accessible states to the cycling temperature via the basic thermodynamic fluctuation relation
\begin{equation}
\Delta\omega_{ir}\sim \Delta\epsilon = \sqrt{k_BC_v(T_c)T_c^2}
\end{equation}
We argue that this fluctuation is felt by the chromophore and gives rise to the irreversible broadening of the hole. $C_v$ is the specific heat of the protein under the specific conditions (denatured, deuterated, special solvents) at the cycling temperature. The above relation yields a temperature power law behaviour with exponents roughly between 1 and 2. The curves fitted to our experimental results (Fig.3) are based on a superposition of such a temperature power law and a step function assuming a rather narrow Gaussian distribution of barriers.\\
Concerning the observed deuteration effect, we stress that there are many well protected hydrogen in the protein interior which are not subjected to a quick H/D exchange when the deuterated protein is dissolved in a protonated matrix. Deuteration of these hydrogen bonds may change their strength and, accordingly, may have an influence on the energetics of the protein (i.e. on barrier heights, structural energies, density of states). This may influence the specific heat and may give rise to the observed deuteration effect according to the above relation.\\
As to the nature of the step, we can only speculate. An intriguing idea is to associate this step with uniform surface TLS, for which a narrow distribution of tunneling amplitude has been assumed. A narrow tunneling amplitude distribution is also consistent with a narrow distribution of the corresponding barrier heights. Note that the fact that the step is observed for the denatured protein as well, does not contradict this reasoning because on the time scale of the cycling experiment even the denatured protein follows the anomalous power law diffusion.

\section{SUMMARY AND CONCLUSIONS}
In the context of the spectral phenomena discussed in this paper solutions of globular proteins in glassy solvents can be viewed as a dispersion of mesoscopic molecular objects with very specific intrinsic as well as interface features. These specific features are reflected in characteristic features of their low temperature dynamics. We presented experiments on relaxation (aging) as well as on fluctuation phenomena in SD-experiments. We discussed the influence of the solvent on protein dynamics and, in addition, we performed thermal cycling experiments of spectral holes. The main experimental results are the power law in waiting time with an exponent of about 1/4, the power law in the aging behavior with a rather small exponent of 0.07, the strong solvent effects in going from a wet (Gl/W) to a dry (TH) solvent and the peculiar behavior of the denatured protein. We presented various models, namely the model of a \textit{"random walk on a random path"} and the model of \textit{"independent low energy surface excitations"} to account for the observed facts. This latter model is microscopic and can account for many of the features observed in a qualitative way. The present structure of this model, however, is not yet protein specific, but is rather specific to the mesoscopic nature of proteins via the specific surface interactions. However, the surface of proteins, the hydration shell with weakly bound water molecules, with remarkable density variations, is very specific in itself so that many assumptions in the model, especially the large number of independent low energy excitations, may be true for globular proteins.

\section*{ACKNOWLEDGMENTS}
J.F. and V.V.P. acknowledge support by the DFG (SFB 533, B5) and by the Fonds der Chemischen Industrie. A.B. work is supported by TAMS GL fund (account no. 211043) through the Tulane University.


\begin{thebibliography}{20}
\bibitem{Frauen_1998} H. Frauenfelder, F. Parak and R. D. Young, {\it Annu. Rev. Biophys. Biophys. Chem.} {\bf 17}, 451 (1998).
\bibitem{Frauen_1994} H. Frauenfelder and P.G. Wolynes, {\it Physics Today} {\bf 47}, 58 (1994). 
\bibitem{Dill_1997} K.A. Dill and H.S. Chan, {\it Nat. Struct. Biol.} {\bf 4}, 10 (1997).
\bibitem{Frauen_1979} H. Frauenfelder, G.A. Pestsko and D. Tsernoglou, {\it Nature} {\bf 280}, 558 (1979).
\bibitem{Hartmann_1982} H.Hartmann, F. Parak, W. Steigemann, G.A. Petsko, D.R. Ponzi and H. Frauenfelder, {\it Proc. Natl. Acad. Sci. USA} {\bf 70}, 4967 (1982).
\bibitem{Frauen_1991} H. Frauenfelder, S.G. Sligar and P.G. Wolynes, {\it Science} {\bf 254}, 1598 (1991).
\bibitem{Berlin_1995} Y. A. Berlin, S. F. Fischer, N. I. Chekunaev and V. I. Goldanskii, {\it Chem. Phys.} {\bf 200}, 369 (1995).
\bibitem{Hofmann_2003} C. Hofmann, H. Michel, T.J. Aartsma and J. Koehler, {\it Proc. Nat. Acad. Sci. USA} {\bf 100}, 15534 (2003).
\bibitem{Leeson_1997} D.Thorn - Leeson, D.A. Wiersma, K.D. Fritsch and J. Friedrich, {\it J. Phys. Chem. B} {\bf 101}, 6331 (1997).
\bibitem{Garcia_1997} A.E. Garcia, R. Blumenfeld, G. Hummer and J.A. Krumhansl, {\it Physica D} {\bf 107}, 225 (1997).
\bibitem{Becker_1997} O.M. Becker and M. Karplus, {\it J. Chem. Phys.} {\bf 106}, 1495 (1997).
\bibitem{Iben_1989} E.T. Iben, D. Braunstein, W. Doster, H. Frauenfelder, M.K. Hong, J.B. Johnson, S. Luck, P. Ormos, A. Schulte, D.J. Steinbach, A.H. Xie and P.D. Young, {\it Phys. Rev. Lett.} {\bf 62}, 1916 (1989).
\bibitem{Fried_1986} J. Friedrich and D. Haarer, in {\it Optical Spectroscopy of Glasses}, I. Zschokke, eds. (D. Reidel Publishing Company, Dordrecht, 1986)
\bibitem{Koehler_1989} W. Koehler, J. Zollfrank and J. Friedrich, {\it Phys. Rev. B} {\bf 39}, 5414 (1989).
\bibitem{Fritsch_1997} K.D. Fritsch and J. Friedrich, {\it Physica D} 107, 218 (1997).
\bibitem{Zollfrank_1991} J. Zollfrank, J. Friedrich, J.M. Vanderkooi and J. Fidy, {\it J. Chem. Phys.} {\bf 95}, 3134 (1991).
\bibitem{Vanderko_1975} J.M. Vanderkooi and M. Erecinska, {\it Eur. J. Biochem.} {\bf 60}, 199 (1975).
\bibitem{Skinner_1999} J.L. Skinner, J. Friedrich and J. Schlichter, {\it J. Phys. Chem. A} {\bf 103}, 2310 (1999).
\bibitem{Reinecke_1979} T.L. Reinecke, {\it Solid State Commun.} {\bf 32}, 1103 (1979).
\bibitem{Kharl_2002} B.M. Kharlamov and G. Zumofen, {\it J. Chem. Phys.} {\bf 116}, 5107 (2002).
\bibitem{Kador_1993} L. Kador, {\it J. Luminescence} {\bf 56}, 165 (1993).
\bibitem{Koehler_1996} M. Koehler, J. Gafert, J. Friedrich, J.M. Vanderkooi and M. Laberge, {\it Biophys. J.} {\bf 71}, 77 (1996).
\bibitem{Breinl_1984} W. Breinl, J. Friedrich and D. Haarer, {\it J. Chem. Phys.} {\bf 81}, 3915 (1984).
\bibitem{Littau_1991} K.A. Littau and M.D. Fayer, {\it Chem. Phys. Lett.} {\bf 176}, 551 (1991).
\bibitem{Schlich_2001} J. Schlichter and J. Friedrich, {\it J. Chem. Phys.} {\bf 114}, 8718 (2001).
\bibitem{VVP_2004} V.V. Ponkratov, J. Friedrich, D. Markovic, H. Scheer and J.M. Vanderkooi, {\it J. Phys. Chem. B} {\bf 108}, 1109 (2004).
\bibitem{Schlich_2000} J. Schlichter, J. Friedrich, L. Herenyi and J. Fidy, {\it J. Chem. Phys.} {\bf 112}, 3045 (2000).
\bibitem{VVP_2002} V.V. Ponkratov, J. Friedrich and J.M. Vanderkooi, {\it J. Chem. Phys.} {\bf 117}, 4594 (2002).
\bibitem{Koehler_1989_a} W. Koehler and J. Friedrich, {\it J. Chem. Phys.} {\bf 90}, 1270 (1989).
\bibitem{Singh_1984} G.P. Singh, H.J. Schink, H.v. Loehneysen, F. Parak and S. Hunklinger, {\it Z. Phys. B} {\bf 55}, 23 (1984).
\bibitem{Yang_1986} I.S. Yang and A.C. Anderson, {\it Phys. Rev. B} {\bf 34}, 2942 (1986).
\bibitem{Schlich_2001_a} J. Schlichter and J. Friedrich, M. Parbel and H. Scheer, {\it Pho. Sci. News} {\bf 6}, 100 (2001).
\bibitem{Schlich_2001_b} J. Schlichter, J. Friedrich, M. Parbel and H. Scheer, {\it J. Chem. Phys.} {\bf 114}, 9638 (2001).
\bibitem{Schlich_2003} J. Schlichter, V.V. Ponkratov and J. Friedrich, {\it Fizika Nizkih Temp.} {\bf 29}, 1054 (2003).
\bibitem{Schlich_2001_c} J. Schlichter, J. Friedrich, L. Herenyi and J. Fidy, {\it Biophys. J.} {\bf 80}, 2011 (2001).
\bibitem{Schlich_2002} J. Schlichter, J. Friedrich, L. Herenyi and J. Fidy, {\it J. Phys. Chem. B} {\bf 106}, 3510 (2002).
\bibitem{Berlin_1996} Yu.A. Berlin and A.L. Burin, {\it Chem.Phys. Letters} {\bf 257}, 665 (1996).
\bibitem{Berlin_2001} Yu.A. Berlin, A.L. Burin, L.D. Siebbeles and M.A. Ratner, {\it J. Chem. Phys. A} {\bf 105}, 5666 (2001).
\bibitem{Burin_2002} A.L. Burin, Yu. A. Berlin, A.Z. Patashinski, M. A. Ratner and J. Friedrich, {\it Physica B} {\bf 316-317}, 321 (2002).
\bibitem{Hartog_1999} F.T.H. den Hartog, C. van Papendrecht, U. Stoerkel and S. Voelker, {\it J.Phys. Chem. B} {\bf 103}, 1375 (1999).
\bibitem{Vitkup_2000} D. Vitkup, D. Ringe, G. Petsko and M. Karplus, {\it Nat. Str. Biol.} {\bf 7}, 34 (2000).
\bibitem{Doster_1989} W. Doster, S. Cusack and W. Petry, {\it Nature} {\bf 337}, 754 (1989).
\bibitem{Fritsch_1996} K. Fritsch, J. Friedrich and B.M. Kharlamov, {\it J. Chem. Phys.} {\bf 105}, 1798 (1996).
\bibitem{Salvino_1994} D. J. Salvino, S. Rogge, B. Tigner and D. D. Osheroff, {\it Phys. Rev. Lett.} {\bf 73}, 268 (1994).
\bibitem{Burin_1995} A. L. Burin, {\it J. Low Temp. Phys.} {\bf 100}, 309 (1995).
\bibitem{Burin_1998} A. L. Burin, D. Natelson, D. D. Osheroff and Yu. Kagan, in {\it Tunneling Systems in Amorphous Solids, Chapter 5}, P. Esquinazi, eds. (Springer, 1998) 
\bibitem{AHW_P_1972} P. W. Anderson, B. I. Halperin and C. M. Warma, {\it Philos. Mag.} {\bf 25}, 1 (1972); W. A. Phillips, {\it J. Low Temp. Phys.} {\bf 7}, 351 (1972).
\bibitem{BH_1977} J. L. Black and B. I. Halperin, {\it Phys. Rev. B} {\bf 16}, 2879 (1977).
\bibitem{HR_1986} S. Hunklinger and A. K. Raychaudhary, {\it Progr. Low Temp. Phys.} {\bf 9}, 267 (1986).
\bibitem{Berlin_1997} Y. A. Berlin, A. L. Burin and S. F. Fischer, {\it Chem. Phys.} {\bf 220}, 25 (1997).
\bibitem{Neu_1997} P. Neu and A. Heuer, {\it J. Chem. Phys.} {\bf 107}, 8686 (1997).
\bibitem{Patash} A. Z. Patashinski, private communication.
\bibitem{Kagan} Y. Kagan and L. A. Maksimov, {\it Zh. Eksp. Teor. Fiz.} {\bf 84}, 792 (1983); Y. Kagan and L. A. Maksimov, {\it Zh. Eksp. Teor. Fiz.} {\bf 79}, 1363 (1980).
\bibitem{Kokshenev} V. B. Kokshenev, {\it J. Low Temp. Phys.} {\bf 20}, 373 (1975).
\bibitem{Tupitsyn} I. S. Tupitsyn, N. V. Prokofev and P. C. E. Stamp, {\it Int. J. Mod. Phys. B} {\bf 11}, 2901 (1997).
\bibitem{Berlin_1996_a} Yu. A. Berlin, A. L. Burin and V. I. Goldanskii, {\it Z. Phys.} {\bf 37}, 333 (1996).
\bibitem{Li} X. Li, D. Clarkson and H. Meyer, {\it J. Low Temp. Phys.} {\bf 78}, 335 (1990).
\bibitem{Wunder_1998} R. Wunderlich, H. Maier, D. Haarer and B. M. Kharlamov, {\it J. Phys. Chem. B} {\bf 102}, 10150 (1998).
\bibitem{Nittke} A. Nittke, P. Esquinazi, H. C. Semmelhack, A. L. Burin and A. Z. Patashinskii, {\it Eur. Phys. J. B} {\bf 8}, 19 (1999).
\bibitem{Nittke_a} A. Nittke, P. Esquinazi and A. L. Burin, {\it Phys. Rev. B} {\bf 58}, 5374 (1998).
\bibitem{Watson} S. K. Watson and R. O. Pohl, {\it Phys. Rev. B} {\bf 68}, 104203 (2003).
\bibitem{Prokofev} N. V. Prokofev, I. S. Tupitsyn and A. L. Burin, unpublished.
\bibitem{Haarer} P. Neu, R. J. Silbey, A. Heuer, S. J. Zilker and D. Haarer, {\it J. Luminesc.} {\bf 76}, 619 (1998); H. Maier, B. M. Kharlamov and D. Haarer, {\it Phys. Rev. Lett.} {\bf 76}, 2085 (1996).
\bibitem{Scharnagl} C. Scharnagl, to be published
\bibitem{Vanderko} B. Zelent, A.D. Kaposi, N.V. Nucci, K.A. Sharp, S.D. Dalosto, W.W. Wrihgt and J. M. Vanderkooi, to be published.
\end{thebibliography}
\end{document}